\documentclass[final,3p,times]{elsarticle}

\usepackage{amssymb}
\usepackage{amsmath}
\usepackage[cal=cm,scr=euler]{mathalpha}
\usepackage{hyperref}

\usepackage{calc,atbegshi,picture}
\newcommand{\placepreprint}{%
\AtBeginShipoutFirst{%
\put(
\textwidth+\oddsidemargin-\widthof{ MIT-CTP/6085, CERN-TH-2026-193},
-2pt-\topmargin-\heightof{ MIT-CTP/6085, CERN-TH-2026-193}
){\normalfont MIT-CTP/6085, CERN-TH-2026-193}
}
}

\journal{Journal of Subatomic Particles and Cosmology}

\begin{document}

\placepreprint

\begin{frontmatter}



\title{Heavy quark transport in quark-gluon plasma beyond the non-relativistic limit}

\author[mit]{Krishna Rajagopal}
\author[kitp]{Bruno Scheihing-Hitschfeld*}
\author[cern]{Urs Achim Wiedemann}

\affiliation[mit]{organization={Center for Theoretical Physics – a Leinweber Institute},
             addressline={Massachusetts Institute of Technology},
             city={Cambridge},
             state={MA},
             postcode={02139},
             country={USA}}

\affiliation[kitp]{organization={Kavli Institute for Theoretical Physics},
             addressline={University of California},
             city={Santa Barbara},
             state={CA},
             postcode={93106},
             country={USA --- *Speaker at SQM2026}}

\affiliation[cern]{organization={Theoretical Physics Department, CERN},        
             postcode={CH-1211},
             city={Genève 23},
             country={Switzerland}}

\begin{abstract}
The dynamics of heavy quarks in quark-gluon plasma (QGP) formed in heavy ion collisions provide a unique window to characterize its properties. Existing approaches to describe heavy quarks in medium rely either on quasiparticle-based models of QGP, or on assuming that the momentum transfer from the medium follows Gaussian statistics. However, neither of these assumptions can be taken for granted in QCD. In fact, in the prototypical theory that is strongly coupled $\mathcal{N}=4$ SYM, it has been long known there are no momentum-carrying quasiparticles, and furthermore, we have recently shown that the momentum transfer from the medium is far from being Gaussian~\cite{Rajagopal:2025ukd}. Since then, we showed that the asymmetry of said momentum transfer between energy loss and energy gain — which affects all moments of the distribution, not only its Gaussian characteristics — is, in fact, universal~\cite{Rajagopal:2025rxr}.

Therefore, in order to connect the initial heavy quark production cross section with the final hadron  spectrum in a way that is consistent with QFT principles, new methods are needed. In this talk, we present a new transport description of heavy quarks that encodes all of the non-Gaussian features of the momentum transfer from the medium, all of which can be defined and in principle calculated in QCD, without relying on any assumptions regarding the strength of the coupling. As a demonstrative example, we discuss heavy quark equilibration in $\mathcal{N}=4$ SYM. This paves the way towards extracting novel information about fundamental properties of QGP.
\end{abstract}




\end{frontmatter}



\section{Introduction}
\label{sec:intro}

Characterizing the dynamics of quark-gluon plasma (QGP) formed in heavy-ion collisions is a unique opportunity to develop of our understanding of the dynamics of deconfined nuclear matter. Heavy quarks provide one way of probing this phase of matter, as they are formed in the early stages of heavy ion collisions and witness the whole evolution of QGP before decaying into lighter particles. As such, they can be tagged and their dynamics compared between different collision events. The comparison can then be used to constrain the strength and properties of the interaction between QGP and the heavy quark, and of QGP itself.

However, while the interaction mechanism is well understood, in the sense that the field theory Lagrangian that gives rise to the interaction between heavy quarks and the light degrees of freedom of QCD is known, the fact that QGP is a strongly coupled quantum system prevents traditional (perturbative) field theory methods from being directly applicable. Therefore, being able to make structural, nonperturbative statements about the dynamics of the interaction, either quantitative or qualitative, is of great help in order to be able to interpret experimental measurements of heavy flavor suppression in terms of concrete quantities that can be formulated in the underlying field theory.

\section{Heavy Quarks as Brownian Particles: successes and shortcomings of the Langevin equation}

The realization that heavy quarks could be thought of as Brownian particles propagating in a thermal environment provided such a structural organizing principle. Early on in the theoretical study of heavy flavor production in heavy ion collisions~\cite{BrownianLangevinHQ,Moore:2004tg}, 
it was proposed that the non-relativistic limit of heavy quark kinetic transport can be described in terms of a Langevin equation
\begin{equation}
    \frac{dp_i }{dt} = - \eta_D p_i + \xi_i \, ,
\end{equation}
where $p_i$ is the momentum of the heavy quark, $\eta_D$ is the so-called ``drag coefficient,'' and $\xi_i$ is a Gaussian stochastic force characterized by a momentum diffusion coefficient $\kappa$. In the large mass limit, the requirement that this dynamics leads to kinetic equilibrium leads to the Einstein relation 
\begin{equation}
    \kappa = 2 M T \eta_D \label{eq:fluct-diss}
\end{equation}
between momentum fluctuations and energy loss for a heavy quark. Here, $M$ is the mass of the heavy quark and $T$ is the temperature of the surrounding medium. In fact, this relation is a consequence of the fluctuation-dissipation theorem~\cite{Kubo1966}. Based on this, the determination of the heavy quark momentum diffusion coefficient $\kappa$ in QCD has been the subject of substantial effort. The current state of the art of such calculations includes perturbative 
results~\cite{kappaHQ} and non-perturbative lattice calculations~\cite{lattice}, 
painting a satisfactory picture across all temperatures above the crossover~\cite{HotQCD:2025fbd}.

\begin{figure}
    \centering
    \includegraphics[width=0.49\linewidth]{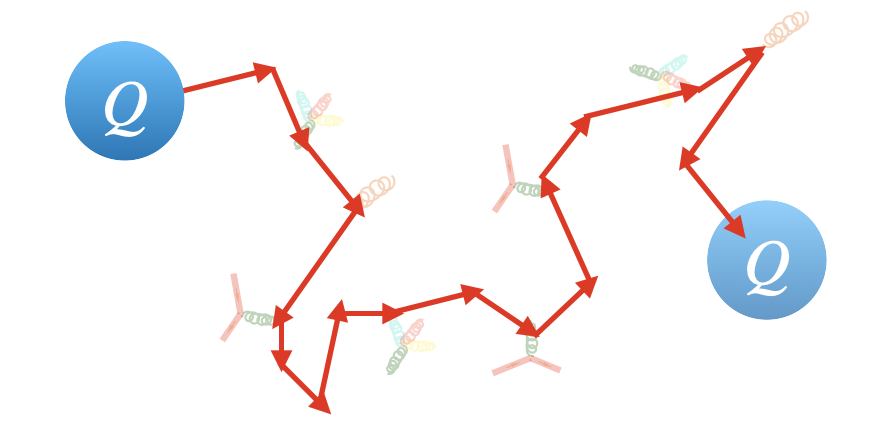}
    \includegraphics[width=0.49\linewidth]{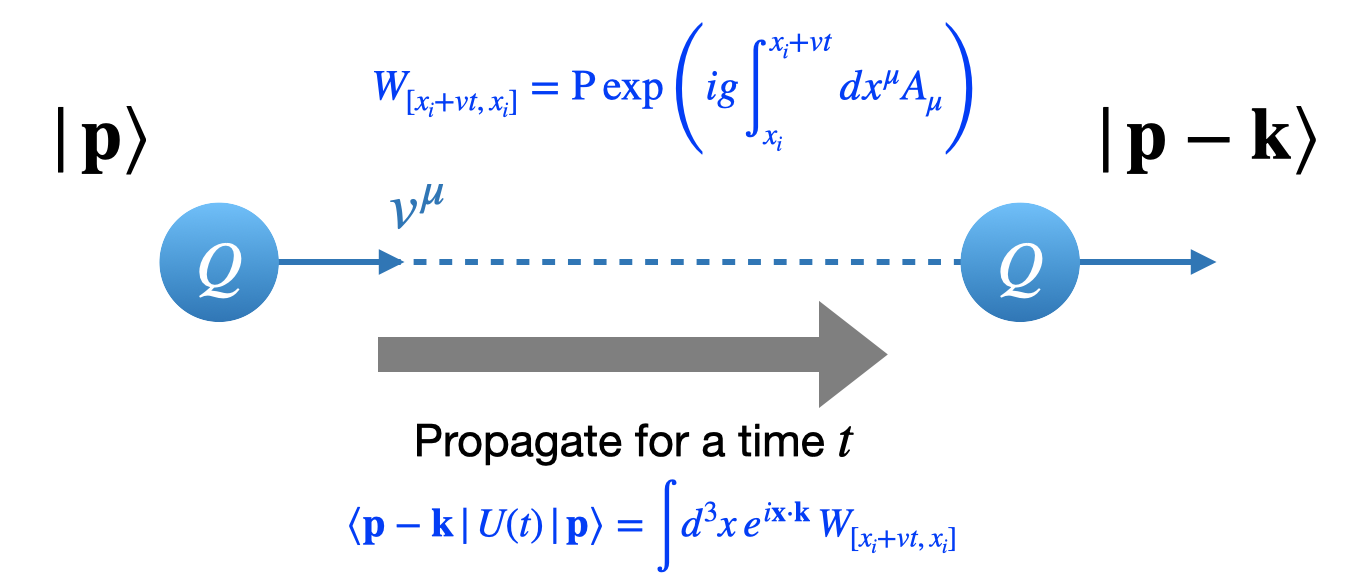}
    \caption{Left: Illustration of the Brownian motion model of a heavy quark in QGP, with randomly distributed momentum kicks governed by a Langevin equation. Right: Quantum mechanical picture of each individual time step, where the transition amplitude between two momentum states of the heavy quark is given by the Fourier transform of a Wilson line along the path of propagation of the heavy quark (a straight line characterized by the heavy quark 4-velocity $v^\mu$).}
    \label{fig:HQ-illustrations}
\end{figure}

However, away from the non-relativistic limit much less is known. While the drag and diffusion coefficients for a moving heavy quark (i.e., with a Lorentz boost factor $\gamma = (1-v^2)^{-1/2} > 1$ in the rest frame of the QGP) are known at weak coupling~\cite{Moore:2004tg}, the Einstein relation
\begin{equation}
    \kappa_L = 2 M \gamma T \eta_D \label{eq:einstein}
\end{equation}
between the longitudinal momentum diffusion coefficient and the drag coefficient, which would ensure kinetic equilibration in a Langevin description in the same way as Eq.~\eqref{eq:fluct-diss} does for the non-relativistic case, only holds at leading logarithm in the QCD coupling constant in perturbation theory, and fails at higher order. This is not an accident, but a feature of these transport coefficients: early calculations in strongly coupled $\mathcal{N}=4$ SYM theory showed that the discrepancy can be quite large~\cite{AdSCFT}. 
More recently, it has been shown that the breaking (already remarked upon in~\cite{Moore:2004tg}) is actually substantial at weak coupling already beyond logarithmic accuracy~\cite{DuPlessis:2026pyr}. 

Given that most heavy quark suppression data in heavy-ion collisions comes from heavy quarks that are either mildly or very relativistic, the foregoing implies that understanding the equilibration process of heavy quarks requires going beyond the assumptions of the Langevin picture.

\section{A Universal Equilibration Condition}

Recently, we showed that the resolution to the breaking of the Einstein relation does not come from considering a qualitatively different model, but rather from a careful accounting of all of the non-Gaussian features of the momentum transfer to the heavy quark~\cite{Rajagopal:2025rxr}. Instead of assuming that the stochastic process that governs heavy quark transport is Gaussian, we consider the momentum change probability as implied by the underlying field theory. In the heavy quark limit (at leading order in a $1/M$ expansion), it is given by the Fourier transform of a Wilson loop
\begin{align}
    P({\bf k};{\bf v}) &=  \int \frac{d^3{\bf L}}{(2\pi)^3}  \, e^{-i {\bf k} \cdot {\bf L}  } \frac{1}{Z} {\rm Tr}_{\mathcal{H}} \! \left[ W^{lk}_{[(0, {\bf L}),(t, {\bf v} t + {\bf L} )]} \tilde{\rho}_{ki} W^{ij}_{[(t, {\bf v} t ),(0,{\bf 0})]} e^{-\beta H}  \rho_{jl} \right] \equiv  \int \frac{d^3{\bf L}}{(2\pi)^3} \, e^{-i {\bf k} \cdot {\bf L}  } \langle W_{\bf v} \rangle ({\bf L}) \, , \label{eq:P-from-W}
\end{align}
where ${\bf k}$ is the momentum change, defined as positive for energy loss, ${\bf v}$ is the 3-velocity of the heavy quark, $\rho$ and $\tilde{\rho}$ encode information about the heavy quark initial state preparation and final state measurement, respectively, $e^{-\beta H}$ prepares the thermal state of the QGP, $Z$ is a normalization constant, and 
\begin{equation}
    W^{ij}_{[x_f,x_i]} = P \exp \left( i g \int_{t_i}^{t_f} dt \,  A_{\mu} \dot{x}^\mu \right) \, , \label{eq:wilson-line}
\end{equation}
is the expression for the Wilson lines in terms of the gauge field.

This momentum change probability $P({\bf k})$ defines a stochastic process for the heavy quark, which can be written in terms of a Kolmogorov equation for the momentum space distribution of heavy quarks
\begin{align}
    \partial_\tau \mathscr{P} = - T K(\partial_{\bf p}, {\bf p}) \mathscr{P} \, ,
    \label{eq:Kolmogorov}
\end{align}
where the time evolution operator of this equation is directly determined by the Wilson loop
\begin{equation}
    K({\bf x};{\bf p}) = - \lim_{t \to \infty} \frac{1}{tT} \log \left[ \langle W_{{\bf v}({\bf p})} \rangle({\bf L} = - i {\bf x}) \right] \, .
\end{equation}
This process is Markovian on account of the separation of scales $M \gg T$, which implies that a long time (relative to $1/T$) needs to pass before the momentum of the heavy quark gets appreciably modified. As such, memory of the dynamics of the previous time steps is lost because the correlation time of the environment is $\sim 1/T$.

In~\cite{Rajagopal:2025rxr}, we showed that the Wilson loop $\langle W_{\bf v} \rangle ({\bf L})$ defines a stochastic process that has the Boltzmann distribution with weights determined by the relativistic dispersion relation $E({\bf p}) = \sqrt{{\bf p}^2 + M^2}$ of the heavy quark as a stationary distribution. That is to say, if said Wilson loop is calculated in any field theory, it is guaranteed to have an equilibrium distribution consistent with general principles in statistical physics. Concretely, in~\cite{Rajagopal:2025rxr} we proved an equilibrium condition that can be stated either in terms of the evolution operator $K({\bf x};{\bf p})$ or the momentum change probability $P({\bf k};{\bf v})$:
\begin{equation}
    K({\bf x}; {\bf p}) = K(- {\bf x} - {\bf v}({\bf p})/T ; {\bf p}) \iff P({\bf k};{\bf v}) = P(-{\bf k}; {\bf v}) \exp \left( {\bf v} \cdot {\bf k}/T \right) \, .
\end{equation}
Either of these expressions, which are equivalent for a Markovian process, encodes the detailed balance condition that the stochastic process satisfies around the Boltzmann distribution. This is, simply put, a universal equilibration condition for heavy quarks.

\begin{figure}
    \centering
    \includegraphics[width=0.39\linewidth]{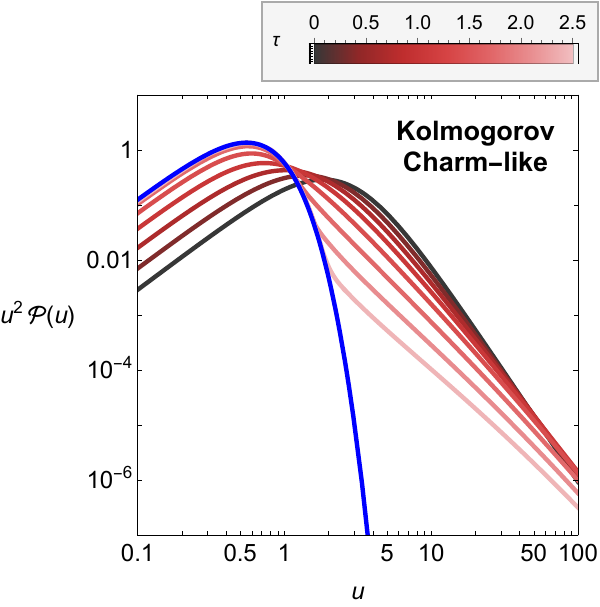}
    \includegraphics[width=0.39\linewidth]{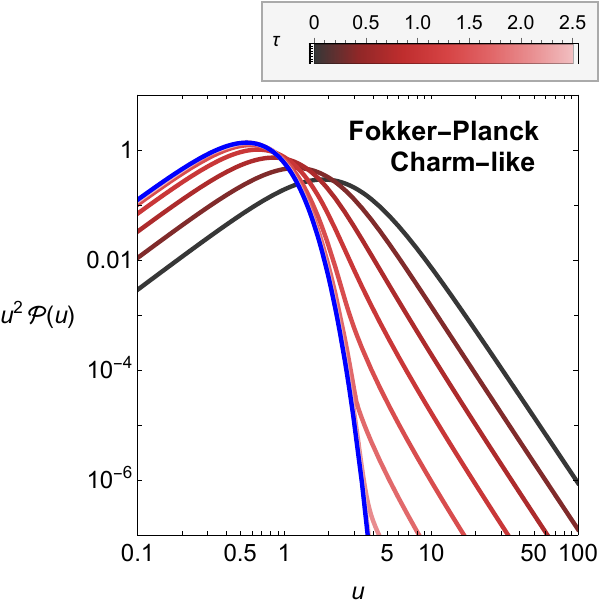}
    \caption{Left: evolution of the heavy quark momentum distribution following the full Kolmogorov dynamics in $\mathcal{N}=4$ SYM theory. Right: evolution of the heavy quark momentum distribution following Fokker-Planck (Gaussian) dynamics, with the same drag coefficient but with the diffusion coefficient chosen by hand to satisfy the Einstein relation. The black curve is the initial condition (same for both) and the blue curve is a Boltzmann distribution. Time and momentum have been rescaled as $\tau = t \sqrt{\lambda} T^2/M$ and $u = p/M$, where $\lambda = g^2 N_c$ is the 't Hooft coupling. The ratio of mass to temperature for these figures is $M/T = 1.5/0.2$, comparable to that of a charm quark in a heavy ion collision. Figure from~\cite{Rajagopal:2026urq}.}
    \label{fig:example-evolution}
\end{figure}

\section{A Concrete Example in Strongly Coupled $\mathcal{N} = 4$ SYM and a Look Ahead}

While the equilibration condition is guaranteed to hold in any field theory, this condition doesn't tell us by itself whether a Gaussian stochastic process is a good approximation to the dynamics or not. For that, one needs to do a calculation. However, in hindsight, we know that the effect of non-Gaussian features must be significant at relativistic velocities because of the failure of the Einstein relation~\eqref{eq:einstein} --- otherwise, heavy quarks would not equilibrate.

Based on the results for $P({\bf k})$ from~\cite{Rajagopal:2025ukd}, we showed in a recent study~\cite{Rajagopal:2026urq} that the effect of non-gaussianities in the momentum transfer probability can be sizable, and leads to an increased survival probability at high momentum relative to what would happen if one constructed a Gaussian stochastic process with the same mean (drag force) but with a variance (diffusion coefficient) chosen by hand so as to ensure that the Einstein relation is satisfied. See Figure~\ref{fig:example-evolution} for a comparison of the evolution of a heavy quark distribution in an infinite $\mathcal{N}=4$ thermal medium with a fixed temperature with and without non-Gaussian features.

Going forward, these results motivate a systematic study of non-Gaussian features of the heavy quark momentum change probability in QCD. A first step was carried out in~\cite{DuPlessis:2026pyr}, where leading order results in perturbation theory were obtained for all moments of the longitudinal momentum transfer distribution for the first time. It would be interesting to extend this computation to next-to-leading order, and to either calculate or constrain moments of the distribution via non-perturbative lattice QCD techniques.
Given that the beyond-Gaussian moments cannot be neglected in the prototypical theory that is $\mathcal{N}=4$ SYM, where they are crucial to the process of equilibration both at weak coupling and at strong coupling, advancing the state of the art for constraining the moments of the momentum change probability for heavy quarks with $\gamma > 1$ to the same point as for the non-relativistic limit~\cite{HotQCD:2025fbd} will greatly advance the prospects of interpreting data from heavy ion collisions in terms of QCD from first principles.

\vspace{0.5cm}

{\footnotesize
Research supported in part by grant NSF PHY-2309135 to the Kavli Institute for Theoretical Physics (KITP); by the U.S.~Department of Energy, Office of Science, Office of Nuclear Physics under grant Contract Number DE-SC0011090; and by grant 994312 from the Simons Foundation.}




\end{document}